\documentclass{article}
\usepackage[utf8]{inputenc}
\usepackage{oz2tex}

\begin{document}

\title{Ideas for the future of Prolog inspired by Oz}
\author{Peter Van Roy \\
Universit\'e catholique de Louvain
\and 
Seif Haridi \\
Royal Institute of Technology}
\date{September 16, 2022}

\maketitle

\begin{abstract}
Both Prolog and Oz are multiparadigm languages with a 
logic programming core.
There is a significant subset of Oz that is a syntactic variant of Prolog:
pure Prolog programs with green or blue cuts and \verb+bagof/3+
or \verb+setof/3+ can be translated directly to Oz.
Because of this close relationship between Prolog and Oz, we propose
that the extensions made by Oz to logic programming can be an inspiration
for the future evolution of Prolog.
We explain three extensions, namely
deterministic logic programming,
lazy concurrent functional programming,
and purely functional distributed computing.
We briefly present these extensions and we explain
how they can help Prolog evolve in its next 50 years.
\end{abstract}

\section{Introduction}

The Prolog language was invented 50 years ago and today in 2022 it is still widely used.
Many possible extensions were and are being defined for Prolog to allow
it to evolve as a living language.
The Oz language was developed starting in 1991
and it led to a major system release in 1999 called Mozart \cite{hopl}.
The main goal of Oz was to cleanly support many programming paradigms.
The Oz language and Mozart system saw wide use for about one decade,
but then its use declined.
After 2009 and up to the present day, it is mainly used only for programming education.
This is unlike Prolog, which still enjoys wide use up to the present day.

Despite its decline in usage, Oz pioneered many important concepts that
have moved into the mainstream since its release.
We mention deterministic dataflow computation, actors that return futures,
deep support for constraint programming through computation spaces,
and deep embedding of distributed computing.
This is relevant for Prolog because of the
close semantic relationship between Oz and Prolog.
The semantic foundation of Oz is concurrent constraint programming,
which is a generalization of Prolog's execution model \cite{saraswat90}.
Prolog programs with logical semantics can be directly
translated into Oz, as explained in Section \ref{syntax}.
Given these facts, we believe that
the design of Oz can be an inspiration for the future evolution of Prolog,
so that Prolog can enjoy another 50 years of success.
The purpose of this article is to present three important extensions
to Prolog-style logic programming made by Oz and explain why
they can be an inspiration to future Prolog development:
\begin{itemize}
\item {\em Deterministic logic programming} (Section \ref{dlp}).
We show how Oz supports writing logic programs that support
managing deterministic execution while maintaining their logical semantics.
For a given logical semantics, we show how Oz can express different operational
semantics.

\item {\em Lazy concurrent functional programming} (Section \ref{lcfp}).
We show how the logic programming core of Oz can be the foundation of a series
of functional programming paradigms, culminating in lazy deterministic dataflow.
All these paradigms keep the strong confluence properties of pure functional
programming as well as their origin in deterministic logic programming.

\item {\em Purely functional distributed computing} (Section \ref{deep}).
The Oz community built a distributed implementation of Oz based on a deep embedding
approach, where each language entity has its own distributed algorithm.
As part of this work, we designed and proved correctness of distributed
rational tree unification.
Based on the use of logic variables for dataflow synchronization,
we showed that asynchronous message passing can be purely functional.
Finally, we showed that most practical distributed systems used today
greatly overuse nondeterminism, and that it would be better to design
them as mostly functional, adding nondeterminism only in the few places where
it is needed.
\end{itemize}
Each of these extensions is presented separately in its own section,
In each case, we explain the main ideas and we give references for readers
wishing to investigate further.
As a preliminary step, Section \ref{syntax} explains how to translate Prolog
programs into Oz, to justify our claim that Oz is a syntactic variant of Prolog.

\section{Oz is a syntactic variant of Prolog}
\label{syntax}

We explain how to translate any pure Prolog program
extended with green or blue cuts and
the \verb+bagof/3+ or \verb+setof/3+ predicates
into an Oz program with
the same logical and operational semantics as the Prolog program.
We assume the reader has enough knowledge of Prolog to understand our examples.
For brevity, we do not explain the semantics of the Oz statements in detail.
Their meaning should be clear from the examples.
For a more complete and formal explanation of the relationship
between Oz and Prolog, we refer the reader to Chapter 9 of \cite{ctm}.

Both Oz and Prolog support symbolic data structures that are bound
using unification.
Similar to many modern Prolog systems, Oz uses rational tree unification.
There are slight differences in syntax and semantics
between Oz and Prolog data structures, mainly because Oz supports additional
data structures such as records.
We do not explain these differences here;
we refer the reader to \cite{ctm} for a precise definition of Oz data structures.

\subsection{Deterministic predicates}
A predicate is deterministic if all its clauses are logically disjoint.
This is the case when the clauses have disjoint guards.
Deterministic predicates are translated using the Oz statements
\OzInline{\OzKeyword{if}} and \OzInline{\OzKeyword{case}}, which both have a logical semantics.
Consider the following deterministic Prolog predicate:
\begin{verbatim}
 place_queens(I, _, _, _) :- I==0, !.
 place_queens(I, Cs, Us, [_|Ds]) :-
    I>0, J is I-1,
    place_queens(J, Cs, [_|Us], Ds),
    place_queen(I, Cs, Us, Ds).
\end{verbatim}
This predicate has a
\index{blue cut (in Prolog)|emph}blue cut according
to \index{O'Keefe, Richard}O'Keefe~\cite{craft},
i.e., the cut is needed 
to inform naive implementations that the
predicate is deterministic, so they can improve efficiency,
but it does not change the program's results.
The predicate \verb+place_queens/4+ is translated into the following Oz code: 
\begin{oz2texdisplay}\OzSpace{1}\OzKeyword{proc}\OzSpace{1}\OzChar\{PlaceQueens\OzSpace{1}N\OzSpace{1}Cs\OzSpace{1}Us\OzSpace{1}Ds\OzChar\}\OzEol
\OzSpace{4}\OzKeyword{if}\OzSpace{1}N==0\OzSpace{1}\OzKeyword{then}\OzSpace{1}\OzKeyword{skip}\OzEol
\OzSpace{4}\OzKeyword{elseif}\OzSpace{1}N>0\OzSpace{1}\OzKeyword{then}\OzSpace{1}Ds2\OzSpace{1}Us2=\OzChar\_|Us\OzSpace{1}\OzKeyword{in}\OzEol
\OzSpace{7}Ds=\OzChar\_|Ds2\OzEol
\OzSpace{7}\OzChar\{PlaceQueens\OzSpace{1}N-1\OzSpace{1}Cs\OzSpace{1}Us2\OzSpace{1}Ds2\OzChar\}\OzEol
\OzSpace{7}\OzChar\{PlaceQueen\OzSpace{1}N\OzSpace{1}Cs\OzSpace{1}Us\OzSpace{1}Ds\OzChar\}\OzEol
\OzSpace{4}\OzKeyword{else}\OzSpace{1}\OzKeyword{fail}\OzSpace{1}\OzKeyword{end}\OzEol
\OzSpace{1}\OzKeyword{end}\end{oz2texdisplay}

\subsection{Nondeterministic predicates}
A predicate is nondeterministic when its clauses are not logically disjoint.
Such predicates can be backtracked into, possibly giving multiple solutions.
Nondeterministic predicates are translated using the Oz statement \OzInline{\OzKeyword{choice}}.
Consider the following nondeterministic Prolog predicate:
\begin{verbatim}
 place_queen(N, [N|_],   [N|_],   [N|_]).
 place_queen(N, [_|Cs2], [_|Us2], [_|Ds2]) :-
    place_queen(N, Cs2, Us2, Ds2).
\end{verbatim}
This predicate is translated into a nondeterministic Oz procedure:
\begin{oz2texdisplay}\OzSpace{1}\OzKeyword{proc}\OzSpace{1}\OzChar\{PlaceQueen\OzSpace{1}N\OzSpace{1}Cs\OzSpace{1}Us\OzSpace{1}Ds\OzChar\}\OzEol
\OzSpace{4}\OzKeyword{choice}\OzSpace{1}N|\OzChar\_\OzSpace{2}=Cs\OzSpace{1}N|\OzChar\_\OzSpace{2}=Us\OzSpace{1}N|\OzChar\_\OzSpace{2}=Ds\OzEol
\OzSpace{4}[]\OzSpace{5}\OzChar\_|Cs2=Cs\OzSpace{1}\OzChar\_|Us2=Us\OzSpace{1}\OzChar\_|Ds2=Ds\OzSpace{1}\OzKeyword{in}\OzEol
\OzSpace{7}\OzChar\{PlaceQueen\OzSpace{1}N\OzSpace{1}Cs2\OzSpace{1}Us2\OzSpace{1}Ds2\OzChar\}\OzEol
\OzSpace{4}\OzKeyword{end}\OzEol
\OzSpace{1}\OzKeyword{end}\end{oz2texdisplay}

\subsection{First-class Prolog top level}
The previous two predicates are part of a program that computes solutions
to the $n$-queens problem: how to place $n$ queens on an $n \times n$ chessboard so that
no queen attacks another.
For completeness, we give the full program here:
\begin{oz2texdisplay}\OzSpace{1}\OzKeyword{fun}\OzSpace{1}\OzChar\{Queens\OzSpace{1}N\OzChar\}\OzEol
\OzSpace{4}\OzKeyword{fun}\OzSpace{1}\OzChar\{MakeList\OzSpace{1}N\OzChar\}\OzEol
\OzSpace{7}\OzKeyword{if}\OzSpace{1}N==0\OzSpace{1}\OzKeyword{then}\OzSpace{1}nil\OzSpace{1}\OzKeyword{else}\OzSpace{1}\OzChar\_|\OzChar\{MakeList\OzSpace{1}N-1\OzChar\}\OzSpace{1}\OzKeyword{end}\OzEol
\OzSpace{4}\OzKeyword{end}\OzEol
\OzSpace{4}Qs=\OzChar\{MakeList\OzSpace{1}N\OzChar\}\OzEol
\OzSpace{4}\OzKeyword{proc}\OzSpace{1}\OzChar\{PlaceQueens\OzSpace{1}N\OzSpace{1}Cs\OzSpace{1}Us\OzSpace{1}Ds\OzChar\}\OzSpace{1}\OzComment{\OzSpace{1}as\OzSpace{1}before\OzSpace{1}}\OzSpace{1}\OzKeyword{end}\OzEol
\OzSpace{4}\OzKeyword{proc}\OzSpace{1}\OzChar\{PlaceQueen\OzSpace{1}N\OzSpace{1}Cs\OzSpace{1}Us\OzSpace{1}Ds\OzChar\}\OzSpace{1}\OzComment{\OzSpace{1}as\OzSpace{1}before\OzSpace{1}}\OzSpace{1}\OzKeyword{end}\OzEol
\OzSpace{1}\OzKeyword{in}\OzEol
\OzSpace{4}\OzChar\{PlaceQueens\OzSpace{1}N\OzSpace{1}Qs\OzSpace{1}\OzChar\_\OzSpace{1}\OzChar\_\OzChar\}\OzEol
\OzSpace{4}Qs\OzEol
\OzSpace{1}\OzKeyword{end}\end{oz2texdisplay}
This program is run by calling the system procedure
\OzInline{SolveOne} which corresponds to
a Prolog top-level query:
\begin{oz2texdisplay}\OzSpace{1}\OzChar\{Browse\OzSpace{1}\OzChar\{SolveOne\OzSpace{1}\OzKeyword{fun}\OzSpace{1}\OzChar\{\OzChar\$\OzChar\}\OzSpace{1}\OzChar\{Queens\OzSpace{1}8\OzChar\}\OzSpace{1}\OzKeyword{end}\OzChar\}\OzChar\}\end{oz2texdisplay}
The syntax \OzInline{\OzKeyword{fun}\OzSpace{1}\OzChar\{\OzChar\$\OzChar\}\OzSpace{1}...\OzSpace{1}\OzKeyword{end}} defines a zero-argument
function that calls the one-argument function \OzInline{Queens}.
The system procedure \OzInline{Browse} displays its argument.
The answer displayed by \OzInline{Browse} is a list giving the first solution:
\begin{oz2texdisplay}\OzSpace{1}[[1\OzSpace{1}7\OzSpace{1}5\OzSpace{1}8\OzSpace{1}2\OzSpace{1}4\OzSpace{1}6\OzSpace{1}3]]\end{oz2texdisplay}
Oz also provides \OzInline{SolveAll} to compute
all solutions and \OzInline{Solve} to compute a lazy list of solutions.
The \OzInline{Solve} corresponds closely to a Prolog top level where solutions
are computed on demand.

\subsection{Predicates with green or blue cuts}

If your Prolog program uses cut ``!'', then the translation to Oz is simple if the
cut is a green or blue cut as defined by O'Keefe \cite{craft}.  A green cut removes
irrelevant solutions.  A blue cut indicates to the
compiler that the program is deterministic.
To show the translation scheme, we translate the following predicate:
\begin{verbatim}
 foo(X, Z) :- guard1(X, Y), !, body1(Y, Z).
 foo(X, Z) :- guard2(X, Y), !, body2(Y, Z).
\end{verbatim}
The two guards must not bind any head variables, i.e., they are quiet guards.
It is good Prolog style to postpone
binding head variables until after the cut.
The translation has two cases, depending on
whether the guards are deterministic or not.
If a guard is deterministic (it has no \OzInline{\OzKeyword{choice}}),
then it can be written as a deterministic boolean function.
This gives the following simple translation:
\begin{oz2texdisplay}\OzSpace{1}\OzKeyword{proc}\OzSpace{1}\OzChar\{Foo\OzSpace{1}X\OzSpace{1}Z\OzChar\}\OzEol
\OzSpace{4}\OzKeyword{if}\OzSpace{5}Y\OzSpace{1}\OzKeyword{in}\OzSpace{1}\OzChar\{Guard1\OzSpace{1}X\OzSpace{1}Y\OzChar\}\OzSpace{1}\OzKeyword{then}\OzSpace{1}\OzChar\{Body1\OzSpace{1}Y\OzSpace{1}Z\OzChar\}\OzEol
\OzSpace{4}\OzKeyword{elseif}\OzSpace{1}Y\OzSpace{1}\OzKeyword{in}\OzSpace{1}\OzChar\{Guard2\OzSpace{1}X\OzSpace{1}Y\OzChar\}\OzSpace{1}\OzKeyword{then}\OzSpace{1}\OzChar\{Body2\OzSpace{1}Y\OzSpace{1}Z\OzChar\}\OzEol
\OzSpace{4}\OzKeyword{else}\OzSpace{1}\OzKeyword{fail}\OzSpace{1}\OzKeyword{end}\OzEol
\OzSpace{1}\OzKeyword{end}\end{oz2texdisplay}
If a guard is nondeterministic (it uses \OzInline{\OzKeyword{choice}}),
then it can be written with one input
and one output argument, like this: \OzInline{\OzChar\{Guard1\OzSpace{1}In\OzSpace{1}Out\OzChar\}}.
It should not bind the input argument.
This gives the following translation:
\begin{oz2texdisplay}\OzSpace{1}\OzKeyword{proc}\OzSpace{1}\OzChar\{Foo\OzSpace{1}X\OzSpace{1}Z\OzChar\}\OzEol
\OzSpace{4}\OzKeyword{case}\OzSpace{1}\OzChar\{SolveOne\OzSpace{1}\OzKeyword{fun}\OzSpace{1}\OzChar\{\OzChar\$\OzChar\}\OzSpace{1}\OzChar\{Guard1\OzSpace{1}X\OzChar\}\OzSpace{1}\OzKeyword{end}\OzChar\}\OzSpace{1}\OzKeyword{of}\OzSpace{1}[Y]\OzSpace{1}\OzKeyword{then}\OzEol
\OzSpace{7}\OzChar\{Body1\OzSpace{1}Y\OzSpace{1}Z\OzChar\}\OzEol
\OzSpace{4}\OzKeyword{elsecase}\OzSpace{1}\OzChar\{SolveOne\OzSpace{1}\OzKeyword{fun}\OzSpace{1}\OzChar\{\OzChar\$\OzChar\}\OzSpace{1}\OzChar\{Guard2\OzSpace{1}X\OzChar\}\OzSpace{1}\OzKeyword{end}\OzChar\}\OzSpace{1}\OzKeyword{of}\OzSpace{1}[Y]\OzSpace{1}\OzKeyword{then}\OzEol
\OzSpace{7}\OzChar\{Body2\OzSpace{1}Y\OzSpace{1}Z\OzChar\}\OzEol
\OzSpace{4}\OzKeyword{else}\OzSpace{1}\OzKeyword{fail}\OzSpace{1}\OzKeyword{then}\OzEol
\OzSpace{1}\OzKeyword{end}\end{oz2texdisplay}
If neither of these two cases apply to your Prolog program, e.g.,
either your guards bind head variables
or you use cuts in other ways (i.e., as red cuts),
then your program likely does not have a logical semantics.
A program with red cuts is defined only by its operational semantics
and this is outside the scope of our translation scheme.

\subsection{The {\tt bagof/3} and {\tt setof/3} predicates}

Prolog's \verb+bagof/3+ predicate
corresponds to using \OzInline{SolveAll} inside an Oz program.
Its extension \verb+setof/3+ sorts the result and removes duplicates.
This can be done with the Oz built-in \OzInline{Sort} operation.
We show how to translate \verb+bagof/3+ both without and with existential quantification.
Consider the following small biblical database
(inspired by~\cite{artofprolog}):
\begin{verbatim}
 father(terach,  abraham).
 father(abraham, isaac).
 father(haran,   milcah).
 father(haran,   yiscah).
\end{verbatim}
This can be written as follows in Oz:
\begin{oz2texdisplay}\OzSpace{1}\OzKeyword{proc}\OzSpace{1}\OzChar\{Father\OzSpace{1}F\OzSpace{1}C\OzChar\}\OzEol
\OzSpace{4}\OzKeyword{choice}\OzSpace{1}F=terach\OzSpace{2}C=abraham\OzEol
\OzSpace{4}[]\OzSpace{5}F=abraham\OzSpace{1}C=isaac\OzEol
\OzSpace{4}[]\OzSpace{5}F=haran\OzSpace{3}C=milcah\OzEol
\OzSpace{4}[]\OzSpace{5}F=haran\OzSpace{3}C=yiscah\OzEol
\OzSpace{4}\OzKeyword{end}\OzEol
\OzSpace{1}\OzKeyword{end}\end{oz2texdisplay}                                                                 
Calling \verb+bagof/3+ without existential quantification, e.g.:
\begin{verbatim}
 children1(X, Kids) :- bagof(K, father(X,K), Kids).
\end{verbatim}
is defined as follows with \OzInline{SolveAll}:
\begin{oz2texdisplay}\OzSpace{1}\OzKeyword{proc}\OzSpace{1}\OzChar\{Children1\OzSpace{1}X\OzSpace{1}Kids\OzChar\}\OzEol
\OzSpace{4}\OzChar\{SolveAll\OzSpace{1}\OzKeyword{proc}\OzSpace{1}\OzChar\{\OzChar\$\OzSpace{1}K\OzChar\}\OzSpace{1}\OzChar\{Father\OzSpace{1}X\OzSpace{1}K\OzChar\}\OzSpace{1}\OzKeyword{end}\OzSpace{1}Kids\OzChar\}\OzEol
\OzSpace{1}\OzKeyword{end}\end{oz2texdisplay}
The \OzInline{Children1} definition is deterministic;
it assumes \OzInline{X} is known and it returns \OzInline{Kids}.
To search over different values of \OzInline{X}
the following definition should be used instead:
\begin{oz2texdisplay}\OzSpace{1}\OzKeyword{proc}\OzSpace{1}\OzChar\{Children1\OzSpace{1}X\OzSpace{1}Kids\OzChar\}\OzEol
\OzSpace{4}\OzChar\{Father\OzSpace{1}X\OzSpace{1}\OzChar\_\OzChar\}\OzEol
\OzSpace{4}\OzChar\{SolveAll\OzSpace{1}\OzKeyword{proc}\OzSpace{1}\OzChar\{\OzChar\$\OzSpace{1}K\OzChar\}\OzSpace{1}\OzChar\{Father\OzSpace{1}X\OzSpace{1}K\OzChar\}\OzSpace{1}\OzKeyword{end}\OzSpace{1}Kids\OzChar\}\OzEol
\OzSpace{1}\OzKeyword{end}\end{oz2texdisplay}
The call \OzInline{\OzChar\{Father\OzSpace{1}X\OzSpace{1}\OzChar\_\OzChar\}} creates a choice point on \OzInline{X}.
The ``\OzInline{\OzChar\_}'' is syntactic sugar for \OzInline{\OzKeyword{local}\OzSpace{1}X\OzSpace{1}\OzKeyword{in}\OzSpace{1}X\OzSpace{1}\OzKeyword{end}},
which is just a new variable with a very small scope.                           

\noindent
Calling \verb+bagof/3+ with existential quantification, e.g.:
\begin{verbatim}
 children2(Kids) :- bagof(K, X^father(X,K), Kids).
\end{verbatim}
is defined as follows with \OzInline{SolveAll}:
\begin{oz2texdisplay}\OzSpace{1}\OzKeyword{proc}\OzSpace{1}\OzChar\{Children2\OzSpace{1}?Kids\OzChar\}\OzEol
\OzSpace{4}\OzChar\{SolveAll\OzSpace{1}\OzKeyword{proc}\OzSpace{1}\OzChar\{\OzChar\$\OzSpace{1}K\OzChar\}\OzSpace{1}\OzChar\{Father\OzSpace{1}\OzChar\_\OzSpace{1}K\OzChar\}\OzSpace{1}\OzKeyword{end}\OzSpace{1}Kids\OzChar\}\OzEol
\OzSpace{1}\OzKeyword{end}\end{oz2texdisplay}
The Oz solution uses \OzInline{\OzChar\_} to add a new existentially scoped variable.
The Prolog solution, on the other hand, introduces
a new syntactic concept, namely the
``existential quantifier'' \verb+X^+,
which only has meaning in terms of \verb+setof/3+ and \verb+bagof/3+.
The fact that this notation denotes an existential quantifier
is defined explicitly in the Prolog semantics.
The Oz solution, on the other hand, requires no new semantics.

In addition to doing all-solutions \verb+bagof/3+,
Oz programs can do a lazy \verb+bagof/3+,
i.e., where each new solution is calculated on demand.
Lazy \verb+bagof/3+ is done by \OzInline{Solve}, which
returns a lazy list of solutions.

\section{Deterministic logic programming}
\label{dlp}

Writing a logic program in Prolog or another logic language consists in defining
the logical semantics and then choosing an operational semantics that gives a
satisfactory efficiency.  
This follows Kowalski's equation ``Algorithm = Logic + Control''.
Logic and control need to be balanced.
The art of logic programming consists in balancing two 
conflicting tensions: the logical semantics should be simple and the operational
semantics should be efficient.  
When done well, this gives an elegant style in Prolog \cite{craft}.
Oz supports this kind of program design in logic programming by supporting both
deterministic and nondeterministic control flow.
For example, in Prolog we can define a list append predicate as follows:
\begin{verbatim}
 append([], L2, L2).
 append([X|M1], L2, [X|M3]) :- append(M1, L2, M3).
\end{verbatim}
This definition follows Prolog's operational semantics.
Compilers can optimize this to make it deterministic in certain cases,
for example if the first argument is bound to a list.
In Oz we can define the operational semantics more precisely.
Let us show three ways that the append can be defined in Oz.

\subsection{Nondeterministic append}

In our first definition, we define
the append nondeterministically:
\begin{oz2texdisplay}\OzSpace{1}\OzKeyword{proc}\OzSpace{1}\OzChar\{Append\OzSpace{1}L1\OzSpace{1}L2\OzSpace{1}L3\OzChar\}\OzEol
\OzSpace{4}\OzKeyword{choice}\OzEol
\OzSpace{7}L1=nil\OzSpace{2}L3=L2\OzEol
\OzSpace{4}[]\OzSpace{1}X\OzSpace{1}M1\OzSpace{1}M3\OzSpace{1}\OzKeyword{in}\OzEol
\OzSpace{7}L1=X|M1\OzSpace{1}L3=X|M3\OzSpace{1}\OzChar\{Append\OzSpace{1}M1\OzSpace{1}L2\OzSpace{1}M3\OzChar\}\OzEol
\OzSpace{4}\OzKeyword{end}\OzEol
\OzSpace{1}\OzKeyword{end}\end{oz2texdisplay}
This has the same operational semantics as Prolog.

\subsection{Deterministic append (first version)}

We give another definition of append that has the same logical semantics
as before but a deterministic operational semantics:
\begin{oz2texdisplay}\OzSpace{1}\OzKeyword{fun}\OzSpace{1}\OzChar\{Append\OzSpace{1}A\OzSpace{1}B\OzChar\}\OzEol
\OzSpace{4}\OzKeyword{case}\OzSpace{1}A\OzEol
\OzSpace{4}\OzKeyword{of}\OzSpace{1}nil\OzSpace{1}\OzKeyword{then}\OzSpace{1}B\OzEol
\OzSpace{4}[]\OzSpace{1}X|As\OzSpace{1}\OzKeyword{then}\OzSpace{1}X|\OzChar\{Append\OzSpace{1}As\OzSpace{1}B\OzChar\}\OzEol
\OzSpace{4}\OzKeyword{end}\OzEol
\OzSpace{1}\OzKeyword{end}\end{oz2texdisplay}
In this case, argument \OzInline{A} is bound to a list so execution is directional.
Oz allows to use a functional syntax for such definitions.

\subsection{Deterministic append (second version)}

We give yet another definition that again has the same logical semantics
but a second deterministic operational semantics:
\begin{oz2texdisplay}\OzSpace{1}\OzKeyword{fun}\OzSpace{1}\OzChar\{Append\OzSpace{1}B\OzSpace{1}C\OzChar\}\OzEol
\OzSpace{4}\OzKeyword{if}\OzSpace{1}B==C\OzSpace{1}\OzKeyword{then}\OzSpace{1}nil\OzEol
\OzSpace{4}\OzKeyword{else}\OzEol
\OzSpace{7}\OzKeyword{case}\OzSpace{1}C\OzSpace{1}\OzKeyword{of}\OzSpace{1}X|Cs\OzSpace{1}\OzKeyword{then}\OzSpace{1}X|\OzChar\{Append\OzSpace{1}B\OzSpace{1}Cs\OzChar\}\OzSpace{1}\OzKeyword{end}\OzEol
\OzSpace{4}\OzKeyword{end}\OzEol
\OzSpace{1}\OzKeyword{end}\end{oz2texdisplay}
This version of \OzInline{Append} takes the last two arguments as inputs
and returns the first argument as its output.
For example, \OzInline{\OzChar\{Append\OzSpace{1}[3]\OzSpace{1}[1\OzSpace{1}2\OzSpace{1}3]\OzChar\}} returns \OzInline{[1\OzSpace{1}2]}.
Correct execution requires that the second argument is a suffix of the third argument.

\section{Lazy concurrent functional programming}
\label{lcfp}

An important insight of the Oz project was that the logic programming core
of Oz can also support functional programming paradigms.
We show this in four steps:
\begin{itemize}
\item {\em Functional programming with values}.
Widely used functional programming languages, such as Scheme or Haskell, compute
with values.
This can be done in Oz simply by not using unbound logic variables.
To support higher-order programming,
we extend the Oz computation model with function values.
Variables can be bound to function values, giving traditional eager functional programming.
\item {\em Functional programming with logic variables}.
We add logic variables to functional programming with values. 
The bind operation is unification.
This gives exactly the deterministic logic programming paradigm of Section \ref{dlp}.
This paradigm has a surprising benefit: many more function definitions become tail-recursive.
\item {\em Deterministic dataflow}.
We add concurrency to functional programming with logic variables.
We allow any statement to execute in its own thread where
threads synchronize on variable binding.
This is exactly the synchronization model of concurrent logic programming.
Compared to concurrent logic programming, however, we are still purely functional.
\item {\em Lazy deterministic dataflow}.
The final extension adds on-demand computation.  We add a new synchronization operation
that waits until another thread waits on a variable being bound.
With this new primitive operation (called \OzInline{WaitNeeded}), we have added all the power
of lazy evaluation to deterministic dataflow.
\end{itemize}
We examine these four steps in more detail.

\subsection{Functional programming with values}

We extend the computation model with function values.
We allow variables to be bound to lexically scoped closures.
This gives pure functional programming with eager evaluation.
Pure functional programming is the foundation of higher-order
programming, which underlies most of the development of data abstraction, such
as object-oriented programming, abstract data types,
components, templates, metaclasses, and so forth.

In addition to being a foundation for data abstraction,
this model has strong formal properties, 
which are different from but analogous to the strong properties of logic programming.  
The main property is known as the Church-Rosser property, often referred to
as confluence, which states that the final result of an expression reduction
is independent of the reduction choices made during the reduction.
In what follows, we show how Oz exploits the synergy that is obtained by combining
confluence with logic variables.

\subsection{Functional programming with logic variables}

We now extend functional programming with logic variables.
This corresponds exactly to the 
deterministic logic programming of Section \ref{dlp},
so this form of functional programming is in fact doing logic programming.
What have we gained from this extension?
First of all, it is now possible to write functional programs
with partially instantiated data structures, just like in Prolog.

But we have gained much more than just the use of partial data structures.
Allowing unbound logic variables makes possible reduction orders that
are impossible in a functional language based on values only.
For  example, the two deterministic append predicates defined in
Section \ref{dlp} are both tail-recursive,
which is not possible in a functional language
based on values without doing complex program transformation.

Let us examine precisely how adding logic variables makes functions tail-recursive.
We start with the first deterministic append from the previous section.
Section \ref{dlp} defines it syntactically as a function,
but in Oz functions are just syntactic sugar for procedures.
The deterministic append is actually defined as a procedure:
\begin{oz2texdisplay}\OzSpace{1}\OzKeyword{proc}\OzSpace{1}\OzChar\{Append\OzSpace{1}A\OzSpace{1}B\OzSpace{1}C\OzChar\}\OzEol
\OzSpace{4}\OzKeyword{case}\OzSpace{1}A\OzEol
\OzSpace{4}\OzKeyword{of}\OzSpace{1}nil\OzSpace{1}\OzKeyword{then}\OzSpace{1}C=B\OzEol
\OzSpace{4}[]\OzSpace{1}X|As\OzSpace{1}\OzKeyword{then}\OzSpace{1}Cs\OzSpace{1}\OzKeyword{in}\OzEol
\OzSpace{7}C=X|Cs\OzEol
\OzSpace{7}\OzChar\{Append\OzSpace{1}As\OzSpace{1}B\OzSpace{1}Cs\OzChar\}\OzEol
\OzSpace{4}\OzKeyword{end}\OzEol
\OzSpace{1}\OzKeyword{end}\end{oz2texdisplay}
Look at the recursive call, \OzInline{\OzChar\{Append\OzSpace{1}As\OzSpace{1}B\OzSpace{1}Cs\OzChar\}}.
This call comes after the binding \OzInline{C=X|Cs} which
incrementally builds the output of the append.
\OzInline{C} is bound to a new cons cell consisting of \OzInline{X} paired with
a new unbound variable \OzInline{Cs}.
The recursive call can be a tail call because we pass it the unbound variable \OzInline{Cs}.
This is not possible in functional programming with values only.

\subsection{Deterministic dataflow}

We extend the previous paradigm by adding threads and using logic variables
for synchronization.
To understand this paradigm, we first explain the dataflow 
behavior of logic variables.  Assume we declare a logic variable \OzInline{X}.
In a first thread, we do \OzInline{X=10}.  In a second thread, we do \OzInline{Y=X+1}.
Because of the scheduler, we do not know which thread will run first.
The second thread, if it is scheduled first, will suspend when it attempts
to do the addition.  It will wait until \OzInline{X} is bound.
When the first thread binds \OzInline{X}, then the second thread becomes runnable again
and the addition will complete, binding \OzInline{Y} to 11.
To summarize, this is a declarative form of dataflow
with two basic operations, namely binding a logic variable and waiting
until the variable is bound.

Dataflow synchronization is such an important use of logic variables
that we call them {\em dataflow variables}
when we explain the history of Oz \cite{hopl}.
The Oz community did not invent this; it was pioneered in concurrent logic programming,
which was an early development of Prolog for concurrent systems \cite{lncs259*30,tr003}.
The Oz contributions to this concept are its use for purely functional
concurrent and distributed computing and its extension to lazy evaluation.

In deterministic dataflow,
any instruction can execute in its own thread.
This lets us define networks of concurrent agents that communicate through streams,
where we define a {\em stream} as a list
that is built incrementally and that may have an unbound tail.
For example, consider the following sequential functional program:
\begin{oz2texdisplay}\OzSpace{1}\OzKeyword{fun}\OzSpace{1}\OzChar\{Gen\OzSpace{1}L\OzSpace{1}H\OzChar\}\OzEol
\OzSpace{4}\OzChar\{Delay\OzSpace{1}1000\OzChar\}\OzSpace{1}\OzEolComment{\OzSpace{1}Suspend\OzSpace{1}thread\OzSpace{1}execution\OzSpace{1}for\OzSpace{1}1000\OzSpace{1}ms}\OzSpace{4}\OzKeyword{if}\OzSpace{1}L>H\OzSpace{1}\OzKeyword{then}\OzSpace{1}nil\OzSpace{1}\OzKeyword{else}\OzSpace{1}L|\OzChar\{Gen\OzSpace{1}L+1\OzSpace{1}H\OzChar\}\OzSpace{1}\OzKeyword{end}\OzEol
\OzSpace{1}\OzKeyword{end}\OzEol
\OzEol
\OzSpace{1}Xs=\OzChar\{Gen\OzSpace{1}1\OzSpace{1}10\OzChar\}\OzEol
\OzSpace{1}Ys=\OzChar\{Map\OzSpace{1}Xs\OzSpace{1}\OzKeyword{fun}\OzSpace{1}\OzChar\{\OzChar\$\OzSpace{1}X\OzChar\}\OzSpace{1}X*X\OzSpace{1}\OzKeyword{end}\OzChar\}\OzEol
\OzSpace{1}\OzChar\{Browse\OzSpace{1}Ys\OzChar\}\end{oz2texdisplay}
This computes a list of successive integers, squares each element of this list,
and displays the result.
To follow the execution during an interactive session,
we have slowed down the generation so it takes
one second per element.
We can make this concurrent by doing
the generation and mapping in their own threads:
\begin{oz2texdisplay}\OzSpace{1}\OzKeyword{thread}\OzSpace{1}Xs=\OzChar\{Gen\OzSpace{1}1\OzSpace{1}10\OzChar\}\OzSpace{1}\OzKeyword{end}\OzEol
\OzSpace{1}\OzKeyword{thread}\OzSpace{1}Ys=\OzChar\{Map\OzSpace{1}Xs\OzSpace{1}\OzKeyword{fun}\OzSpace{1}\OzChar\{\OzChar\$\OzSpace{1}X\OzChar\}\OzSpace{1}X*X\OzSpace{1}\OzKeyword{end}\OzChar\}\OzSpace{1}\OzKeyword{end}\OzEol
\OzSpace{1}\OzChar\{Browse\OzSpace{1}Ys\OzChar\}\end{oz2texdisplay}
This uses the \OzInline{\OzKeyword{thread}} statement to create new threads.
What is the difference between the concurrent and
the sequential versions?
The result of the calculation is the same in both
cases, namely \OzInline{[1\OzSpace{1}4\OzSpace{1}9\OzSpace{1}16\OzSpace{1}...\OzSpace{1}81\OzSpace{1}100]}.
So what is different?
In the sequential version, \OzInline{Gen} calculates
the whole list before \OzInline{Map} starts.
The final result is displayed all at once
when the calculation is complete, which happens after ten seconds.
In the concurrent version, \OzInline{Gen} and \OzInline{Map} both
execute concurrently.
Whenever \OzInline{Gen} adds an element to its list,
\OzInline{Map} will immediately calculate its square
before the next element is added.
The result is displayed incrementally
as the elements are generated,
one element each second.
Concurrency has converted a batch computation into an incremental computation
without changing the functional semantics.

The concurrent executions of \OzInline{Gen} and \OzInline{Map} can be considered as
functional agents, where we define an {\em agent} as a concurrent computation that
reads zero or more input streams and writes zero or more output streams.
Because of logic variables, both \OzInline{Gen} and \OzInline{Map} are tail-recursive.
This means that the agents execute with constant stack size.
This justifies calling these computations ``agents''.
The concurrent execution is memory efficient as well as being purely functional.
This shows clearly the synergy between logic variables and concurrency.

\subsection{Lazy deterministic dataflow}

The final extension adds the ability to do lazy evaluation to deterministic dataflow.
In the previous section,
we explained how logic variables are used for dataflow synchronization.
We now extend this dataflow paradigm to do lazy evaluation.
We do this by adding one new operation, \OzInline{WaitNeeded}, which does by-need synchronization 
on logic variables.
The operation \OzInline{\OzChar\{WaitNeeded\OzSpace{1}X\OzChar\}} suspends the
current thread if \OzInline{X} is unbound and no other thread is waiting for \OzInline{X} to be bound.
Otherwise, if another thread is suspended on \OzInline{X} or if \OzInline{X} is bound,
the operation succeeds.
We say that \OzInline{WaitNeeded} ``waits until \OzInline{X} is needed''.
Adding this single operation to the language lets us fully define lazy evaluation.
Oz introduces a syntax sugar to make this easy for the programmer.
For example, we define the function \OzInline{\OzChar\{Ints\OzSpace{1}N\OzChar\}}
that returns a lazy list of successive integers starting with \OzInline{N}:
\begin{oz2texdisplay}\OzSpace{1}\OzKeyword{fun}\OzSpace{1}lazy\OzSpace{1}\OzChar\{Ints\OzSpace{1}N\OzChar\}\OzEol
\OzSpace{4}N|\OzChar\{Ints\OzSpace{1}N+1\OzChar\}\OzEol
\OzSpace{1}\OzKeyword{end}\end{oz2texdisplay}
The ``\OzInline{lazy}'' annotation means that the function does lazy evaluation.
It is syntactic sugar for the following procedure:
\begin{oz2texdisplay}\OzSpace{1}\OzKeyword{proc}\OzSpace{1}\OzChar\{Ints\OzSpace{1}N\OzSpace{1}R\OzChar\}\OzEol
\OzSpace{4}\OzKeyword{thread}\OzEol
\OzSpace{7}\OzChar\{WaitNeeded\OzSpace{1}R\OzChar\}\OzEol
\OzSpace{7}R=N|\OzChar\{Ints\OzSpace{1}N+1\OzChar\}\OzEol
\OzSpace{4}\OzKeyword{end}\OzEol
\OzSpace{1}\OzKeyword{end}\end{oz2texdisplay}
Calling \OzInline{S=\OzChar\{Ints\OzSpace{1}1\OzChar\}} will suspend in the \OzInline{WaitNeeded} call
until another thread needs the first element of \OzInline{S} to run.  When
this happens, the \OzInline{WaitNeeded} call succeeds and \OzInline{R} is bound to a list
with one new element.
The recursive call of \OzInline{Ints} continues this behavior for the next elements.
We note that the compiler can optimize the above definition by reusing the thread,
to avoid the creation of a new thread for each recursive call.
The combination of lazy evaluation and concurrency has been known at least since 
1977 \cite{kahn77}, but as far as we know,
the definition of lazy evaluation in a dataflow paradigm and
its connection to logic programming are both original with Oz.

\paragraph{Summary} We summarize the paradigms introduced in this section.
We start with a first-order paradigm that computes with values.
We extend this paradigm in four steps.
First, we add function values, i.e., lexically scoped closures,
which gives pure functional programming with eager evaluation.
Second, we add logic variables,
which gives deterministic logic programming as seen in Section \ref{dlp}.
A side benefit is that functions that compute recursive data structures such as lists
become tail-recursive.
Third, we add threads and use
the dataflow behavior of logic variables to do synchronization,
which gives deterministic dataflow.
This supports concurrent networks of functional agents that communicate
through shared lists used as communication channels, which we call streams.
Note that because list functions are tail-recursive, these agents use constant stack space.
Fourth, in the final step, we add by-need synchronization using \OzInline{WaitNeeded}.
This gives a paradigm that supports both lazy evaluation and concurrency,
which we call lazy deterministic dataflow.
All four paradigms keep the strong confluence properties
of pure functional programming, as well as being conservative extensions to a
language supporting deterministic logic programming.
Since Prolog is identical to a subset of Oz, we propose that some form of these extensions
could become part of a future evolution of Prolog.

\section{Purely functional distributed computing}
\label{deep}

The Oz research community realized early on that 
the Oz language design would be a good starting point
for building a distributed programming system.
Because the design cleanly separates immutable, dataflow, and mutable language entities,
it would be possible to use a {\em deep embedding approach},
where each language entity would be implemented with its own distributed algorithm.
The distribution support was part of the Mozart system at its first release in 1999.
This design had many important innovations including the following:
\begin{itemize}
\item {\em Deep embedding}.
Each language concept was implemented with a distributed algorithm.
This means that applications written for one distribution structure
can easily be ported to another distribution structure without changing the source code.
The only differences are failure behavior and timing.  Failure behavior can be handled
in a modular way, inspired by techniques from Erlang \cite{collet,erlang}.
\item {\em Distributed rational tree unification}.
The concurrent examples of Section \ref{lcfp} can all be run with unchanged source
code if distributed across different compute nodes.
Concurrent pipelines become asynchronous distributed pipelines, while maintaining
the semantics of pure functional programming.
A key part of this system is the distributed rational tree unification algorithm.
The first formal specification and proof of distributed unification was done during
this work and published in \cite{toplas99}.
\item {\em Global references}, {\em distributed lexical scoping},
{\em distributed garbage collection}, and {\em migratory state}.
These are some of the other properties of this design.
\end{itemize}
In the rest of this section we will focus on what we consider to be
the most significant discovery that was made in the work on distributed
computing for Oz.  This discovery is intimately tied to the logic programming
origins of Oz and it may open a window of opportunity for Prolog.

\paragraph{Asynchronous message passing can be pure}
The deterministic data\-flow paradigm and its lazy extension
can be implemented in a distributed setting.
A dataflow variable can be read on one compute node and bound on another node.
This does a distributed synchronization, which in its general form is implemented
by distributed unification.
The salient property of this distributed synchronization is that it enjoys all
the strong properties of pure functional programming.
Specifically, it means that {\em asynchronous message passing between compute nodes
can be completely pure}.
This property remains mostly unknown in the distributed computing community.
In our experience as members of this community,
we observe that many often insist that asynchronous message passing
is intrinsically impure, which is false.

\paragraph{Overuse of nondeterminism}
Most of today's large distributed systems do not take advantage of this property.
In fact, the contrary is true: {\em today's systems greatly overuse nondeterminism}.
This is one of the main reasons
why it is still difficult to build and debug such systems.
Large distributed systems, such as used daily by Google, Facebook, Twitter, and so on,
must be continually babysitted by experts.
This is not an inevitable property of such systems.
In our view, it is mainly due to the massive overuse of nondeterminism in the design
of these systems.
Every single library that is a part of such a system and that has its own API, which
is true of most libraries, is a point of nondeterminism.

\paragraph{Purely functional distributed computing}
The problem of overuse of nondeterminism and a possible solution
are explained in a keynote talk given by one of the authors at CodeBEAM 2019,
{\em Why Time is Evil in Distributed Systems and What To Do About It} \cite{evil}.
The solution we propose is to use deterministic dataflow programming
as the default paradigm for distributed systems and to add nondeterminism
only where it is needed and nowhere else.
With this solution, most distributed systems become mostly functional
with a limited use of nondeterminism.
This greatly simplifies their development, maintenance, and evolution.
We observe that this solution is intimately tied to logic programming,
since the basic concept that makes it work is the dataflow variable,
which is simply a logic variable used in a concurrent setting.

\paragraph{Role of logic programming}
We believe that an appropriate extension of Prolog or another logic language
can potentially solve this problem
and play a significant role in distributed computing.
If such an extension is not done by the logic programing community,
then the distributed computing community will
reinvent concepts from logic programming as it solves these problems.
This reinvention has already started with the development of CRDTs (Conflict-Free
Replicated Data Types), which have monotonic properties and can replace consensus
algorithms, which are nondeterministic,
by purely functional operations in many cases \cite{crdt}.

\section{Conclusions}
\label{conclusions}

Prolog has enjoyed a relatively large popularity since its conception in 1972
up to the present day,
unlike Oz, which despite having many innovations at its release in 1999,
has seen a decline since 2009 and is only used today for education.
Oz declined for sociological reasons that have nothing to do with its
innovations; in fact we observe that the innovations are all becoming
adopted in modern programming systems despite the decline of Oz.
As members of both the Prolog and Oz communities,
we see this as an opportunity to help Prolog evolve
in order to maintain its popularity.

This paper presents three innovations of Oz,
namely deterministic logic programming, lazy concurrent functional programming,
and purely functional distributed computing.
We propose that it would be straightforward to add these innovations to Prolog
because the core logic languages of Oz and Prolog are quite similar.
Most of pure Prolog has a direct syntactic translation to Oz.
This means that the hard work of formulating and understanding
these innovations in a logic programming context has already been done.
We hope that the ideas presented in this paper will help the future of Prolog
as well as help Oz regain some recognition as a source of innovation.

\end{document}